\documentclass[runningheads,a4paper]{llncs}
\usepackage{makeidx}
\usepackage{amssymb}
\usepackage{amscd, amsfonts}
\usepackage{verbatim}
\usepackage{listings}
\usepackage{commath}
\usepackage{float}
\usepackage{tikz}
\usepackage{isabelle}
\usepackage{isabellesym}
\usepackage{url}

\usepackage{lmodern}
\usepackage[T1]{fontenc}

\newcommand{\refmap}{\mapsto_r}
\newcommand{\arraymap}{\mapsto_a}
\newcommand{\isato}{\isasymrightarrow}
\newcommand{\isaTo}{\isasymRightarrow}

\newcommand{\isaImp}{\isasymLongrightarrow}

\allowdisplaybreaks[1]

\begin{document}

\title{Efficient verification of imperative programs using auto2}

\author{Bohua Zhan}

\institute{Technical University of Munich}
\maketitle

\begin{abstract}
  Auto2 is a recently introduced prover for the proof assistant
  Isabelle. It is designed to be both highly customizable from within
  Isabelle, and also have a powerful proof search mechanism. In this
  paper, we apply auto2 to the verification of imperative programs. We
  describe the setup of auto2 for both stages of the proof process:
  verification of a functional version of the program, and refining to
  the imperative version using separation logic. As examples, we
  verify several data structures, including red-black trees, interval
  trees, priority queues, and union-find. We also verify several
  algorithms making use of these data structures. These examples show
  that our framework is able to verify complex algorithms efficiently
  and in a modular manner.
\end{abstract}

\section{Introduction}

Verification of imperative programs has been a well-studied
area. While work on separation logic addressed the main theoretical
issues, verification in practice is still a tedious process. Even if
we limit to the case of sequential programs with relatively simple
memory-allocation patterns, verification is still difficult when a lot
of mathematical reasoning is required to justify the underlying
algorithm. Such reasoning can quickly go beyond the ability of
automatic theorem provers. Proof assistants such as Isabelle and Coq
provide an environment in which human users can guide the computer
through the proof. However, such a process today often requires a lot
of low-level reasoning with lists, sets, etc, as well as dealing with
details of separation logic. We believe much work can still be done to
provide more automation in this area, reducing the amount of time and
expertise needed to perform verifications, with the goal of eventually
making verification of complex algorithms a routine process.

The auto2 prover in Isabelle is introduced by the author in
\cite{auto2}. Its approach to automation in proof assistants is
significantly different from the two main existing approaches: tactics
and the use of external automatic theorem provers (as represented by
Sledgehammer in Isabelle). Compared to Sledgehammer, auto2 is highly
customizable: users can set up new reasoning rules and procedures at
any point in the development of a theory (for example, our entire
setup for separation logic is built outside the main auto2
program). It also works directly with higher-order logic and types
available in Isabelle. Compared to tactics, auto2 uses a
saturation-based search mechanism, that is closer to the kind of
search performed in automatic theorem provers, and from experience has
been more powerful and stable than the backtracking approach usual in
the tactics framework.

In this paper, we apply auto2 to the verification of imperative
programs. We limit ourselves to sequential programs with relatively
simple memory-allocation patterns. The algorithms underlying the
programs, however, require substantial reasoning to justify. The
verification process can be roughly divided into two stages: verifying
a functional version of the program, and refining it to an imperative
version using separation logic.

The main contributions of this paper are as follows.\footnote{Code
  available at https://github.com/bzhan/auto2}
\begin{itemize}
\item We discuss the setup of auto2 to provide automation for both
  stages of this process. For the verification of functional programs,
  this means automatically proving simple lemmas involving lists,
  sets, etc. For refining to the imperative program, this means
  handling reasoning with separation logic.
\item Using our setup, we verify several data structures including
  red-black trees, interval trees, priority queues, and union-find. We
  also verify algorithms including Dijkstra's algorithm for shortest
  paths and a line-sweeping algorithm for detecting rectangle
  intersection. These examples demonstrate that using our approach,
  complex algorithms can be verified in a highly efficient and modular
  manner.
\end{itemize}

We now give an outline for the rest of the paper. In Section
\ref{sec:auto2}, we give an overview of the auto2 prover. In Section
\ref{sec:functional}, we discuss our setup of auto2 for verification
of functional programs. In Section \ref{sec:intro_imp}, we review the
Imperative HOL framework in Isabelle and its separation logic, which
we use to describe and verify the imperative programs. In Section
\ref{sec:sepauto}, we discuss our setup of auto2 for reasoning with
separation logic. In Section \ref{sec:examples}, we briefly describe
each of the case studies, showing some statistics and comparison with
existing verifications. Finally, we review related work in Section
\ref{sec:related}, and conclude in Section \ref{sec:conclusion}.

\paragraph{Acknowledgements.}
The author would like to thank Adam Chlipala, Peter Lammich, and
Tobias Nipkow for discussions and feedback during this project, and to
the referees for their helpful comments. For the first half of this
project, the author was at MIT and was supported by NSF Award
No. 1400713. During the second half, the author is at TU Munich, and
is supported by DFG Koselleck grant NI 491/16-1.

\section{Overview of the auto2 prover}
\label{sec:auto2}

The auto2 prover is introduced in \cite{auto2}. In \cite{auto2-fol},
several additional features are described, in an extended application
to formalization of mathematics. In this section, we summarize the
important points relevant to this paper.

Auto2 uses a saturation-based proof search mechanism. At any point
during the search, the prover maintains a list of \emph{items}, which
may be derived facts, terms that appeared in the proof, or some other
information. At the beginning, the statement to be proved is converted
into contradiction form, and its assumptions form the initial
state. The search ends successfully when a contradiction is
derived. In addition to the list of items, the prover also maintains
several additional tables, three of which will be described below.

\subsection{Proof steps}

\emph{Proof steps} are functions that produce new items from existing
ones. During the development of an Isabelle theory, proof steps can be
added or removed at any time. At each iteration of the proof search,
auto2 applies the current list of proof steps to generate new
items. Each new item is given a \emph{score} and inserted into a
priority queue. They are then added to the main list of items at
future iterations in increasing order of score. The score is by
default computed from the size of the proposition (smaller means
higher priority), which can be overriden for individual proof steps.

Adding new proof steps is the main way to set up new functionality for
auto2. Proof steps range from simple ones that apply a single theorem,
to complex functions that implement some proof procedure. Several
proof steps can also work together to implement some proof strategy,
communicating through their input and output items. We will see
examples of all these in Sections \ref{sec:functional} and
\ref{sec:sepauto}.

\subsection{Rewrite table}

Among the tables maintained by auto2, the most important is the
\emph{rewrite table}. The rewrite table keeps track of the list of
currently known (ground) equalities. It offers two main operations:
deciding whether two terms are equivalent, and matching up to known
equalities (E-matching). The latter is the basic matching function
used in auto2: whenever we mention matching in the rest of the paper,
it is assumed to mean E-matching using the rewrite table.

We emphasize that when a new ground equality is derived, auto2 does
\emph{not} use it to rewrite terms in the proof state. Instead, the
equality is inserted into the rewrite table, and \emph{incremental}
matching is performed on relevant items to discover new matches.

\subsection{Property and well-formedness tables}
\label{sec:othertables}

We now discuss two other important tables maintained by auto2: the
property table and the well-formedness table.

Any predicate (constant of type \isa{'a \isaTo\ bool}) can be
registered as a \emph{property} during the development of a
theory. During the proof, the \emph{property table} maintains the list
of properties satisfied by each term appearing in the proof. Common
examples of predicates that we register as properties include
sortedness on lists and invariants satisfied by binary search trees.

For any function, we may register certain conditions on its arguments
as \emph{well-formedness conditions} of that function. Common examples
include the condition \isa{a $\ge$ b} for the term \isa{(a - b)::nat},
and \isa{i < length xs} for the term \isa{xs ! i} (\isa{i}'th element
of the list \isa{xs}). We emphasize that registering well-formedness
conditions is for the automation only, and does not imply any
modification to the logic. During the proof, the \emph{well-formedness
  table} maintains the list of well-formedness conditions that are
known for each term appearing in the proof.

The property and well-formedness tables allow proof steps to quickly
lookup certain assumptions of a theorem. We call assumptions that can
be looked-up in this way \emph{side conditions}. We will see examples
of these in Section \ref{sec:simple_proofstep}, and another important
application of the well-formedness table in Section
\ref{sec:norm_sub}.

\subsection{Case analysis}

The need for case analysis introduces further complexities. New case
analysis is produced by proof steps, usually triggered by the
appearance of certain facts or terms in the proof. We follow a
saturation-based approach to case analysis: the list of cases (called
\emph{boxes}) is maintained as a part of the proof state, and
derivation in all boxes are performed in parallel. More precisely,
every item (and entry in the tables) is assigned to some box,
according to the set of additional assumptions needed to derive that
item. When a contradiction is derived in a box with additional
assumption $P$, the fact $\neg P$ is added to its parent box. The
proof finishes only if a contradiction is derived in the initial box
(with no additional assumptions).

\subsection{Proof scripts}

Auto2 defines its own language of proof scripts, which is similar to,
but independent from the Isar proof language in Isabelle. The main
differences between auto2 and Isar are that auto2 scripts do not
contain names of tactics (all subgoals are proved using auto2), labels
for intermediate goals, or names of previous theorems.

Examples of auto2 scripts are given in Section
\ref{sec:example_script}. We explain the basic commands here (all
commands in auto2 scripts begin with an \isa{@} sign, to distinguish
them from similar Isar commands).

\begin{itemize}
\item \isa{@have P}: prove the intermediate goal $P$. Afterwards, make
  $P$ available in the remainder of the proof block.
\item \isa{@case P}: prove the current goal with additional assumption
  $P$. Afterwards, make $\neg P$ available in the remainder of the
  proof block.
\item \isa{@obtain x where P(x)}: here \isa{x} must be a fresh
  variable. Prove the intermediate goal \isamath{\exists x.\
    P(x)}. Afterwards, create variable $x$ and make fact $P(x)$
  available in the remainder of the proof block.
\item \isa{@with $\dots$ @end}: create a new proof block. That is,
  instead of proving the subgoal in the previous command directly
  using auto2, prove it using the commands between \isa{@with} and
  \isa{@end}.
\item \isa{@induct}, \isa{@prop\_induct}, etc: commands for several
  types of induction. Each type of induction has its own syntax,
  specifying which variable or proposition to apply induction on. We
  omit the details here.
\end{itemize}

\section{Verification of functional programs}
\label{sec:functional}

Proofs of correctness of functional programs involve reasoning in many
different domains, such as arithmetic, lists, sets, maps, etc. The
proof of a single lemma may require results from more than one of
these domains. The design of auto2 allows automation for each of these
domains to be specified separately, as a collection of proof
steps. During the proof, they work together by communicating through
the common list of items and other tables maintained by the prover.

In this section, we discuss our setup of auto2 for verification of
functional programs. It is impossible to describe the entire setup in
detail. Instead, we will give some examples, showing the range of
functionality that can be supported in auto2. At the end of the
section, we give an example showing the strength of the resulting
automation.

We emphasize that the aim here is not to implement complete proof
procedures, or to compete with highly-optimized theory solvers for
efficiency. Instead, we simply aim for the prover to consistently
solve tasks that humans consider to be routine. Since we are in an
interactive setting, we can always ask the user to provide
intermediate goals for more difficult proof tasks.

\subsection{Simple proof steps}
\label{sec:simple_proofstep}

Most of the proof steps added to auto2 apply a single theorem. Such
proof steps can be added easily to auto2 (for example, a forward
reasoning rule can be added by setting the \isa{forward} attribute to
a theorem). We describe some basic examples in this section.

\subsubsection{Forward and backward reasoning}

The most basic kind of proof steps apply a theorem in the forward or
backward direction. For example, the theorem
\[ \isa{sorted (x \# xs) \isaImp\ y $\in$ set xs \isaImp\ x $\le$
  y} \] is added as a \emph{forward} proof step. This proof step looks
for pairs of facts in the form \isa{sorted (x \# xs)} and \isa{y $\in$
  set xs} (using E-matching, same below). For every match, it outputs
the fact \isa{x $\le$ y} as a new item (to be added to the main list
of items at a future iteration).

In contrast, the theorem
\[ \isa{sorted xs \isaImp\ j < length xs \isaImp\ i $\le$ j \isaImp\
  xs ! i $\le$ xs ! j}. \] should be added as a \emph{backward} proof
step. This proof step looks for facts of the form \isa{$\neg$(xs ! i
  $\le$ xs ! j)} (equivalently, goal to prove \isa{xs ! i $\le$ xs !
  j}). For every match, it looks for the assumption \isa{sorted xs} in
the property table, and \isa{j < length xs} in the well-formedness
table (it is the well-formedness condition of the subterm \isa{xs !
  j}). If both side conditions are found, the proof step outputs fact
\isa{$\neg$(i $\le$ j)} (equivalently, goal to prove \isa{i $\le$ j}).

Another type of proof step adds a new fact for any term matching a
certain pattern. For example, for the theorem
\[ \isa{n < length xs \isaImp\ xs ! n $\in$ set xs}, \] the
corresponding proof step looks for terms of the form \isa{xs ! n}.
For every match, it looks for the assumption \isa{n < length xs} in
the well-formedness table, and output \isa{xs ! n $\in$ set xs} if the
assumption is found. This particular setup is chosen because
assumptions of the form \isa{y $\in$ set xs} appears frequently in
practice.

\subsubsection{Rewrite rules}

Rewrite rules form another major class of proof steps. They add new
equalities to the rewrite table, usually after matching the left side
of the equality. As an example, consider the theorem for evaluation of
list update:
\[ \isa{i < length xs \isaImp\ xs[i := x] ! j = (if i = j then x else
  xs ! j)}. \] The corresponding proof step looks for terms of the
form \isa{xs[i := x] ! j}. For every match, it looks for the
assumption \isa{i < length xs} in the well-formedness table (this is
the well-formedness condition of \isa{xs[i := x]}). If the assumption
is found, the proof step outputs the equality. When the equality is
pulled from the priority queue at a later iteration, it is added to
the rewrite table.

For the theorem evaluating the length of list update:
\[ \isa{length (xs[i := x]) = length xs} \] we add a slightly
different proof step: it produces the equality whenever it finds the
term \isa{xs[i := x]}, without waiting for \isa{length (xs[i := x])}
to appear. This can be justified by observing that it is useful to
know the length of any list appearing in the proof, as it is mentioned
in the assumptions of many theorems.

\subsubsection{Generating case analysis}\label{sec:case_analysis}

Another class of proof steps generate case analysis on seeing certain
terms or facts in the proof state. For example, there is a proof step
that looks for terms of the form \isa{if P then b else c}, and creates
case analysis on \isa{P} for every match.

Case analysis may also be created to check well-formedness
conditions. Usually, when we register a well-formedness condition,
auto2 will look for the condition in the list of items during the
proof. However, sometimes it is better to be more proactive, and try
to prove the condition whenever a term of the given form appears. This
is achieved by creating a case analysis with the condition as the goal
(or equivalently, with the negation of the condition as the
assumption).

\subsection{Normalization of natural number expressions}
\label{sec:norm_sub}

In this section, we give an example of a more complex proof step. It
compares expressions on natural numbers by normalizing both sides with
respect to addition and subtraction.

Mathematically, the expression $a-b$ on natural numbers is undefined
if $a<b$. In Isabelle (and many other proof assistants), it is simply
defined to be zero. This means many equalities involving subtraction
on natural numbers that look obvious are in fact invalid. Examples
include $a-b+b=a$, which in Isabelle is false if $a<b$.

This substantially complicates normalization of expressions on natural
numbers involving subtraction. In general, normalization of such an
expression agrees with intuition as long as the expression is
well-formed, in the sense of Section \ref{sec:othertables}. Following
the terminology in \cite[Section 3.3]{auto2-fol}, we say a
\emph{well-formed term} is a term together with a list of theorems
justifying its well-formedness conditions, and a \emph{well-formed
  conversion} is a function that, given a well-formed term, returns an
equality rewriting that term, together with theorems justifying
well-formedness conditions on the right side of the
equality. Well-formed conversions can be composed in the same way as
regular conversions (rewriting procedures). In particular, we can
implement normalization for expressions on natural numbers with
respect to addition and subtraction as a well-formed conversion.

This is in turn used to implement the following proof step. Given any
two terms $s,t$ of type \isa{nat} involving addition and subtraction,
look for their well-formedness conditions in the well-formedness
table. If all well-formedness conditions for subtraction are present,
normalize $s$ and $t$ using the well-formed conversion. If their
normalizations are the same, output the equality $s = t$. Such proof
steps, when combined with proof scripts, allow the user to rapidly
perform arithmetic manipulations.

\subsection{Difference logic on natural numbers}

Difference logic is concerned with propositions of the form $a\le b +
n$, where $n$ is a constant. A collection of such inequalities can be
represented as a directed graph, where nodes are terms and weighted
edges represent inequalities between them. A collection of
inequalities is contradictory if and only if the corresponding graph
contains a negative cycle, which can be determined using the
Bellman-Ford algorithm.

In auto2, we implement difference logic for natural numbers using
special items and proof steps. While less efficient than a graph-based
implementation, it is sufficient for our purposes, and also interacts
better with other proof steps. Each inequality on natural numbers is
represented by an item of type \isa{NAT\_ORDER}, which contains a
triple \isa{<a,b,n>} recording the terms on the two sides and the
difference. The transitivity proof step looks for pairs of items of
the form \isa{<a,b,m>} and \isa{<b,c,n>}, and produces the item
\isa{<a,c,m+n>} for each match. The resolve proof step looks for items
of the form \isa{<a,a,n>}, where \isa{n} is less than zero, and
derives a contradiction for each match.

\subsection{Example}
\label{sec:example_script}

As an example, we show a snippet from the functional part of the
verification of the union-find data structure. Union-find is
implemented on an array \isa{l}, with \isa{l ! i} equal to \isa{i} if
\isa{i} is the root of its component, and the parent of \isa{i} if
otherwise. \isa{rep\_of i} denotes the root of the component
containing \isa{i}. The compress operation is defined as:
\begin{isabelle}
\ \ ufa\_compress l x = l[x := rep\_of l x]
\end{isabelle}

The main properties of \isa{ufa\_compress} are stated and proved using
auto2 as fllows:
\begin{isabelle}
lemma ufa\_compress\_invar: \isanewline
\ \ "ufa\_invar l \isaImp\ x < length l \isaImp\ l' = ufa\_compress l x \isaImp\ ufa\_invar l'"
@proof \isanewline
\ \ @have "$\forall$i<length l'. rep\_of\_dom (l', i) $\wedge$ l' ! i < length l'" @with \isanewline
\ \ \ \ @prop\_induct "ufa\_invar l $\wedge$ i < length l" \isanewline
\ \ @end \isanewline
@qed
\isanewline\isanewline
lemma ufa\_compress\_aux: \isanewline
\ \ "ufa\_invar l \isaImp\ x < length l \isaImp\ l' = ufa\_compress l x \isaImp \isanewline
\ \ \ i < length l' \isaImp\ rep\_of l' i = rep\_of l i" \isanewline
@proof @prop\_induct "ufa\_invar l $\wedge$ i < length l" @qed \isanewline

lemma ufa\_compress\_correct: \isanewline
\ \ "ufa\_invar l \isaImp\ x < length l \isaImp\ ufa\_$\alpha$ (ufa\_compress l x) = ufa\_$\alpha$ l" \isanewline
\ \ by auto2
\end{isabelle}

The only hints that needs to be provided by the human to prove these
lemmas are how to apply the induction (specified using the
\isa{@prop\_induct} command). By comparison, in the AFP library
\cite{afp-sep}, the corresponding proofs require 20 tactic invocations
in 42 lines of Isar text.

\section{Imperative HOL and its separation logic}
\label{sec:intro_imp}

In this section, we review some basic concepts from the Imperative HOL
framework in Isabelle and its separation logic. See
\cite{imphol,refinement,afp-sep} for details.

\subsection{Heaps and programs}

In Imperative HOL, procedures are represented as Haskell-style
monads. They operate on a heap (type \isa{heap}) consisting of a
finite mapping from addresses (natural numbers) to natural numbers,
and a finite mapping from addresses to lists of natural numbers (in
order to support arrays). Values of any type \isa{'a} can be stored in
the heap as long as one can specify an injection from \isa{'a} to the
natural numbers. This means records with multiple fields, such as
nodes of a search tree, can be stored at a single address. Along with
native support for arrays, this eliminates any need for pointer
arithmetic.

The type of a procedure returning a value of type \isa{'a} is given by
\[ \isa{datatype 'a Heap = Heap "heap \isaTo\ ('a $\times$ heap)
  option"} \] The procedure takes as input a heap $h$, and outputs
either \isa{None} for failure, or \isa{Some $(r,h')$}, where $r$ is
the return value and $h'$ is the new heap. The \isa{bind} function for
sequencing two procedures has type
\[ \isa{'a Heap \isaTo\ ('a \isaTo\ 'b Heap) \isaTo\ 'b Heap}. \]

Imperative HOL does not have native support for while loops. Instead,
basic applications use recursion throughout, with properties of
recursive procedures proved by induction. We will follow this approach
in our examples.

\subsection{Assertions and Hoare triples}

The type \emph{partial heap} is defined by \isa{pheap = heap $\times$
  nat set}. The partial heap $(h, as)$ represents the part of the heap
$h$ given by the set of addresses $as$.

An assertion (type \isa{assn}) is a mapping from \isa{pheap} to
\isa{bool}, that does not depend on values of the heap outside the
address set. The notation $(h, as) \vDash P$ means ``the assertion $P$
holds on the partial heap $(h, as)$''.

Some basic examples of assertions are:
\begin{itemize}
\item \isa{true}: holds for all valid partial heaps.
\item \isa{emp}: the partial heap is empty.
\item $\uparrow(b)$: the partial heap is empty and $b$ (a boolean
  value) holds.
\item $p\refmap a$: the partial heap contains a single address
  pointing to value $a$.
\item $p\arraymap xs$: the partial heap contains a single address
  pointing to list $xs$.
\end{itemize}

The \emph{separating conjunction} on two assertions is defined as
follows:
\[ P * Q = \lambda(h, as).\, \exists u\;v.\,u \cup v = as \wedge u
\cap v = \emptyset \wedge (h, u)\vDash P \wedge (h, v)\vDash Q. \]
This operation is associative and commutative, with unit
\isa{emp}. Existential quantification on assertions is defined as:
\[ \exists_A x.\, P(x) = \lambda(h, as).\, \exists x.\, (h, as) \vDash
P(x). \]

Assertions of the form $\uparrow(b)$ are called \emph{pure}
assertions. In \cite{afp-sep}, conjunction, disjunction, and the magic
wand operator on assertions are also defined, but we will not use them
here.

A Hoare triple is a predicate of type
\[ \isa{assn \isaTo\ 'a Heap \isaTo\ ('a \isaTo\ assn) \isaTo\
  bool}, \] defined as follows: \isa{<$P$> $c$ <$Q$>} holds if for any
partial heap $(h,as)$ satisfying $P$, the execution of $c$ on $(h,
as)$ is successful with new heap $h'$ and return value $r$, and the
new partial heap $(h',as')$ satisfies $Q(r)$, where $as'$ is $as$
together with the newly allocated addresses.

From these definitions we can prove the Hoare triples for the basic
commands, as well as the \emph{frame rule}
\[ \isa{<$P$> $c$ <$Q$> \isaImp\ <$P*R$> $c$ <$\lambda x.\,
  Q(x)*R$>}. \]

In \cite{afp-sep}, there is further setup of a tactic \isa{sep\_auto}
implementing some level of automation in separation logic. We do not
make use of this tactic in our work.

\section{Automation for separation logic}
\label{sec:sepauto}

In this section, we discuss our setup of auto2 for separation
logic. The setup consists of a collection of proof steps working with
Hoare triples and entailments, implemented in around 2,000 lines of ML
code (including specialized matching for assertions). While knowledge
of auto2 is necessary to implement the setup, we aim to provide an
easy-to-understand interface, so that no knowledge of the internals of
auto2, or of details of separation logic, is needed to use it for
concrete applications.

\subsection{Basic approach}

Our basic approach is to analyze an imperative program in the forward
direction: starting at the first command and finishing at the last,
using existing Hoare triples to analyze each line of the procedure.
To simplify the discussion, suppose the procedure to be verified
consists of a sequence of commands $c_1; \dots; c_n$. Let $P_0$ be the
(spatial) precondition of the Hoare triple to be proved.

To reason about the procedure, we use existing Hoare triples for $c_1,
\dots, c_n$ (these may include the induction hypothesis, if some of
$c_i$ are recursive calls). We write each Hoare triple in the
following standard form:

\begin{isabelle}
<$p_1$ * $\cdots$ * $p_m$ * $\uparrow(a_1)$ * $\cdots$ * $\uparrow(a_k)$>
\isanewline
\ \ c \isanewline
<$\lambda r.\ \exists_A \vec{x}.\ q_1$ * $\cdots$ * $q_n$ *
 $\uparrow(b_1)$ * $\cdots$ * $\uparrow(b_l)$>
\end{isabelle}
Here \isa{$p_1$ * $\cdots$ * $p_m$} is the \emph{spatial} part of the
precondition, specifying the shape of the heap before the command, and
\isa{$\uparrow(a_1)$ * $\cdots$ * $\uparrow(a_k)$} is the \emph{pure}
part of the precondition, specifying additional constraints on the
abstract values (we assume that all variables appearing in $a_i$ also
appear in $p_i$ or $c$). The assertions \isa{$q_1$ * $\cdots$ * $q_n$}
and \isa{$\uparrow(b_1)$ * $\cdots$ * $\uparrow(b_l)$} (depending on
the return value $r$ and possibly new data-variables $\vec{x}$) are
the spatial and pure parts of the postcondition. They provide
information about the shape of the heap after the command, and
constraints on abstract values on that heap.

Applying the Hoare triple for $c_1$ involves the following steps:
\begin{enumerate}
\item Match the pattern $c$ with the command $c_1$, instantiating some
  of the arbitrary variables in the Hoare triple.
\item Match the spatial part of the precondition with $P_0$. This is
  the \emph{frame-inference} step: the matching is up to the
  associative-commutative property of separating conjunction, and only
  a subset of factors in $P_0$ need to be matched. Each match should
  instantiate all remaining arbitrary variables in the Hoare triple.
\item Generate case analysis (discussed at the end of Section
  \ref{sec:case_analysis}) to try to prove each of the pure conditions
  $a_i$.
\item After all pure conditions are proved, apply the Hoare
  triple. This creates new variables for the return value $r$ and
  possible data variables $\vec{x}$. The procedure is replaced by
  $c_2;\dots;c_n$ and the precondition is replaced by \isa{$q_1$ *
    $\cdots$ * $q_n$}. The pure assertions $b_1,\dots,b_l$ in the
  postcondition are outputed as facts.
\end{enumerate}

On reaching the end of the imperative program, the goal reduces to an
entailment, which is solved using similar matching schemes as above.

\subsection{Inductively-defined assertions}
\label{sec:ind_assn}

Certain assertions, such as those for linked lists and binary trees,
are defined inductively. For example, the assertion for binary trees
(with a key-value pair at each node) is defined as follows:

\begin{isabelle}
btree Tip p = $\uparrow$(p = None) \isanewline
btree (tree.Node lt k v rt) (Some p) = \isanewline
\ \ \ \ \ \ ($\exists_A$lp rp. p $\refmap$ Node lp k v rp * btree lt lp * btree rt rp) \isanewline
btree (tree.Node lt k v rt) None = false
\end{isabelle}

Here \isa{btree $t$ $p$} is an assertion stating that the memory
location $p$ contains a functional data structure $t$. The term
\isa{tree.Node lt k v rt} represents a functional binary tree, where
\isa{lt} and \isa{rt} are subtrees, while the term \isa{Node lp k v
  rp} represents a record on the heap, where \isa{lp} and \isa{rp} are
pointers. When working with inductively-defined assertions like this,
the heap can be divided into spatial components in several ways. For
example, a heap satisfying the assertion
\begin{equation}\label{eq:ind_assn1}
  \isa{p $\refmap$ Node lp k v rp * btree lt lp * btree rt
    rp}
\end{equation}
also satisfies the assertion
\begin{equation}\label{eq:ind_assn2}
  \isa{btree (tree.Node lt k v rt) p}.
\end{equation}
The former considers the heap as three components, while the latter
considers it as one component.

We follow the policy of always using assertions in the more expanded
form (that is, (\ref{eq:ind_assn1}) instead of
(\ref{eq:ind_assn2})). This means matching of assertions must also
take into account inductive definitions of assertions, so that the
assertion (\ref{eq:ind_assn1}) will match the pattern \isa{btree ?t p
  * ?P} as well as (for example) the pattern \isa{btree ?t lp *
  ?P}. This is realized by maintaining a list of inductive definitions
of assertions in the theory, and have the special matching function
for assertions refer to this list during matching.

\subsection{Modularity}

For any data structure, there are usually two levels at which we can
define assertions: the concrete level with definition by induction or
in terms of simpler data structures, and the abstract level describing
what data the structure is supposed to represent.

For example, in the case of binary trees, the concrete assertion
\isa{btree} is defined in the previous section. At the abstract level,
a binary tree represents a mapping. The corresponding assertion
\isa{btree\_map} is defined by:
\begin{isabelle}
btree\_map M p = ($\exists_A$t. btree t p * $\uparrow$(tree\_sorted t) * $\uparrow$(M = tree\_map t)),
\end{isabelle}
where \isa{tree\_map $t$} is the mapping corresponding to the binary
tree $t$ with key-value pairs at each node. For each operation on
binary trees, we first prove a Hoare triple on the concrete assertion
\isa{btree}, then use it to derive a second Hoare triple on the
abstract assertion \isa{btree\_map}. For example, for the insertion
operation, we first show:
\begin{isabelle}
<btree t b> btree\_insert k v b <btree (tree\_insert k v t)>
\end{isabelle}
where \isa{tree\_insert} is the functional version of insertion on
binary trees. Using this Hoare triple, and the fact that
\isa{tree\_insert} preserves sortedness and behaves correctly with
respect to \isa{tree\_map}, we prove
\begin{isabelle}
<btree\_map M b> btree\_insert k v b <btree\_map (M \{k $\isato$ v\})>
\end{isabelle}

Similarly, for tree search, the Hoare triple on the concrete assertion
is:
\begin{isabelle}
<btree t b * $\uparrow$(tree\_sorted t)> \isanewline
btree\_search x b \isanewline
<$\lambda$r. btree t b * $\uparrow$(r = tree\_search t x)>
\end{isabelle}
This Hoare triple, along with properties of \isa{tree\_search}, is
used to prove the Hoare triple on the abstract assertion:
\begin{isabelle}
<btree\_map M b> btree\_search x b <$\lambda$r. btree\_map M b * $\uparrow$(r = M$\langle$x$\rangle$)>"
\end{isabelle}

After the Hoare triples for \isa{btree\_map} are proved, the
definition of \isa{btree\_map}, as well as the Hoare triples for
\isa{btree}, can be hidden from auto2 by removing the corresponding
proof steps. This enforces modularity of proofs: auto2 will only use
Hoare triples for \isa{btree\_map} from now on, without looking into
the internal implementation of the binary tree.

\subsection{Example}

With the above setup for separation logic, auto2 is able to prove the
correctness of the imperative version of compression in union-find
after specifying how to apply induction (using the \isa{@prop\_induct}
command):

\begin{isabelle}
uf\_compress i ci p = ( \isanewline
\ \ if i = ci then return () \isanewline
\ \ else do \{ \isanewline
\ \ \ \ ni $\leftarrow$ Array.nth p i; \isanewline
\ \ \ \ uf\_compress ni ci p; \isanewline
\ \ \ \ Array.upd i ci p; \isanewline
\ \ \ \ return () \isanewline
\ \ \}) \isanewline
\isanewline
lemma uf\_compress\_rule: \isanewline
\ \ "ufa\_invar l \isaImp\ i < length l \isaImp\ ci = rep\_of l i \isaImp \isanewline
\ \ \ <p $\arraymap$ l> \isanewline
\ \ \ uf\_compress i ci p \isanewline
\ \ \ <$\lambda$\_. $\exists_A$l'. p $\arraymap$ l' * $\uparrow$(ufa\_invar l' $\wedge$ length l' = length l $\wedge$ \isanewline
\ \ \ \ \ \ \ \ \ \ \ \ \ ($\forall$i<length l. rep\_of l' i = rep\_of l i))>" \isanewline
@proof @prop\_induct "ufa\_invar l $\wedge$ i < length l" @qed
\end{isabelle}

Note that the imperative procedure performs full compression along a
path, rather than a single compression for the functional version in
Section \ref{sec:example_script}. By comparison, the corresponding
proof in the AFP requires 13 tactic invocations (including 4
invocations of \isa{sep\_auto}) in 34 lines of Isar text.

\section{Case studies}
\label{sec:examples}

In this section, we describe the main case studies performed to
validate our framework. For each case study, we describe the data
structure or algorithm that is being verified, its main difficulties,
and then give comparisons to existing work. Statistics for the case
studies are summarized in the following table. On a laptop with two
2.0GHz cores and 16GB of RAM, it takes auto2 approximately 14 minutes
to process all of the examples. \newline \newline
\begin{tabular} {c|c|c|c|c|c|c}
  & \#Imp & \#Def & \#Thm & \#Step & Ratio & \#LOC \\ \hline
  Union-find & 49 & 7 & 26 & 42 & 0.86 & 244 \\
  Red-black tree & 270 & 27 & 83 & 173 & 0.64 & 998 \\
  Interval tree & 84 & 17 & 50 & 83 & 0.99 & 520 \\
  Rectangle intersection & 33 & 18 & 31 & 111 & 3.36 & 417 \\
  Indexed priority queue & 83 & 10 & 53 & 84 & 1.01 & 477 \\
  Dijkstra's algorithm & 44 & 19 & 62 & 150 & 3.41 & 549
\end{tabular}
\newline \newline
The meaning of the fields are as follows:
\begin{itemize}
\item \#Imp is the number of lines of imperative code to be verified.
\item \#Def is the number of definitions made during the verification
  (not counting definitions of imperative procedures).
\item \#Thm is the number of lemmas and theorems proved during the
  verification.
\item \#Step is the number of ``steps'' in the proof. Each definition,
  lemma, and intermediate goal in the proof script counts as one step
  (so for example, a lemma proved with one intermediate goal counts as
  two steps). We only count steps where auto2 does some work, omitting
  for example variable definitions.
\item Ratio: ratio between \#Step and \#Imp, serving as a measure of
  the overhead of verification.
\item \#LOC: total number of lines of code in the theories
  (verification of functional and imperative program). This can be
  used to make approximate comparisons with other work.
\end{itemize}

\subsection{Union-find}

Our verification follows closely that in the AFP \cite{afp-sep}. As in
the example in the AFP, we do not verify that the array containing the
size of components has reasonable values (important only for
performance analysis). Two snippets of auto2 proofs are shown in
previous sections. Overall, we reduced the number of lines in the
theory by roughly a half. In a further example, we applied union-find
to verify an algorithm for determining connectivity on undirected
graphs (not counted in the statistics).

\subsection{Red-black tree}

We verified the functional red-black tree given by Okasaki
(\cite{okasaki}, for insertion) and Kahrs (\cite{kahrs}, for
deletion). Both functional correctness and maintenance of invariants
are proved. We then verified an imperative version of the same
algorithm (imperative in the sense that no more memory is allocated
than necessary). This offers a stringent test for matching involving
inductively defined assertions (discussed in Section
\ref{sec:ind_assn}).  For the functional part of the proof, we used
the technique introduced by Nipkow \cite{nipkow-funtree} for proving
sortedness and proper behavior on the associated maps using the
inorder traversal as an intermediary.

Functional red-black tree has been verified several times in proof
assistants \cite{appel-rbt,nipkow-funtree}. The imperative version is
a common test-case for verification using automatic theorem provers
\cite{chin,qiu1,qiu2,grasshopper}. It is also verified
``auto-actively'' in the SPARK system \cite{spark}, but apparently not
in proof assistants such as Coq and Isabelle.

\subsection{Interval tree and rectangle intersection}

Interval tree is an augmented data structure, with some version of
binary search tree serving as the base. It represents a set of
intervals $S$, and offers the operation of determining whether a given
interval $i$ intersects any of the intervals in $S$. See \cite[Section
14.3]{clrs} for details. For simplicity, we verified interval tree
based on an ordinary binary search tree.

As an application of interval trees, we verify an algorithm for
detecting rectangle intersection (see \cite[Exercise
14.3-7]{clrs}). Given a collection $S$ of rectangles aligned to the
$x$ and $y$ axes, one can determine whether there exists two
rectangles in $S$ that intersect each other using a line-sweeping
algorithm as follows. For each rectangle $[a,b]\times [c,d]$, we
create two operations: adding the interval $[a,b]$ at time $c$, and
removing it at time $d$. The operations for all rectangles are sorted
by time (breaking ties by putting insertion before deletion) and
applied to an initially empty interval tree. There is an intersection
if and only if at some point, we try to insert an interval which
intersects an existing interval in the tree. Formal verification of
interval trees and the line-sweeping algorithm for rectangle
intersection appear to be new.

\subsection{Indexed priority queue and Dijkstra's algorithm}

The usual priority queue is implemented on one array. It supports
insertion and deleting the minimum. In order to support decreasing the
value of a key (necessary for Dijkstra's algorithm), we need one more
``index'' array recording locations of keys. Having two arrays produce
additional difficulty in having to verify that they stay in sync in
all operations.

The indexed priority queue is applied to verify a basic version of
Dijkstra's algorithm. We make several simplifying assumptions: the
vertices of the graph are natural numbers from 0 to $n-1$, and there
is exactly one directed edge between each ordered pair of vertices, so
that the weights of the graph can be represented as a matrix. Since
the matrix is unchanged during the proof, we also do not put it on the
heap. Nevertheless, the verification, starting from the definition of
graphs and paths, contains all the essential ideas of Dijkstra's
algorithm.

The indexed priority queue and Dijkstra's algorithm are previously
verified using the refinement framework in
\cite{refinement2,dijkstra}. It is difficult to make precise
comparisons, since the approach used in the refinement framework is
quite different, and Dijkstra's algorithm is verified there without
the above simplifying assumptions. By a pure line count, our
formalization is about 2-3 times shorter.

\section{Related work}
\label{sec:related}

This paper is a continuation of the work in \cite{auto2} and
\cite{auto2-fol}. There is already some verification of imperative
programs in \cite{auto2}. However, they do not make use of separation
logic, and the examples are quite basic. In this paper, we make full
use of separation logic and present more advanced examples.

The refinement framework, introduced by Lammich in \cite{refinement},
can also be used to verify programs in Imperative-HOL. It applies
refinement and data abstraction formally, verifying algorithms by
step-wise refinement from specifications to concrete
implementations. It has been used to verify several advanced graph
algorithms \cite{graph1,graph2,graph3}. Our work is independent from
the refinement framework. In particular, we use refinement and data
abstraction only in an ad-hoc manner.

Outside Imperative-HOL, there are many other frameworks based on
tactics for automating separation logic in proof assistants. Examples
include
\cite{appel-tac,xyfeng,chargueraud,bedrock,ynot,afp-sep,tuerk}. As
discussed in the introduction, our framework is implemented on top of
the auto2 prover, which follows a quite different approach to
automation compared to tactics.

Finally, there are many systems for program verification using
automatic theorem provers. The main examples include
\cite{why3,dafny,verifast}. The basic approach is to generate
verification conditions from user-supplied annotations, and solve them
using SMT-based provers. Compared to such systems, we enjoy the usual
advantages of working in an interactive theorem prover, including a
small trusted kernel, better interaction when proving more difficult
theorems, and having available a large library of mathematical
results.

\section{Conclusion}
\label{sec:conclusion}

In this paper, we described the setup of the auto2 prover to provide
automation for verification of imperative programs. This include both
the verification of a functional version of the program, and refining
it to the imperative version using separation logic. Using our
framework, we verified several data structures and algorithms,
culmulating in Dijkstra's shortest paths algorithm and the
line-sweeping algorithm for detecting rectangle intersection. The case
studies demonstrate that auto2 is able to provide a great deal of
automation in both stages of the verification process, significantly
reducing the length and complexity of the proof scripts required.

\end{document}